\documentclass[11pt]{revtex4} 
\usepackage{graphicx}
\usepackage{epsfig}
\begin{document}

\begin{center}
{\LARGE Non-Maxwellian velocity distribution and anomalous
diffusion of {\it in vitro} kidney cells} \vspace{1cm}

{\large L. Diambra$^{1,*}$, L. C. Cintra$^1$, D. Schubert$^2$, and
L. da F. Costa$^1$} \vspace{1cm}

$^1$Instituto de F\'{\i}sica de S\~ao Carlos, Universidade de
S\~ao Paulo \\ Caixa Postal: 369, cep: 13560-970, S\~ao Carlos SP,
Brazil. \vspace{1cm}

$^2$The Salk Institute, \\ 10010 N. Torrey Pines Road, La Jolla,
CA 92037, USA.
\end{center}


\begin{center}
{\bf Abstract}
\end{center}

This manuscript uses a statistical mechanical approach to study
the effect of the adhesion, through MOCA protein, on cell
locomotion. The MOCA protein regulates cell-cell adhesion, and we
explore its potential role in the cell movement. We present a
series of statistical descriptions of the motion in order to
characterize the cell movement, and found that MOCA affects the
statistical scenario of cell locomotion. In particular, we observe
that MOCA enhances the tendency of joint motion, inhibits
super-diffusion, and decreases overall cell motion. These facts
are compatible with the hypothesis that the cells move faster in a
less cohesive environment. Furthermore, we observe that velocity
distribution tails are longer than those predicted by
Maxwell-Boltzmann in all cases studied here, indicating that cell
movement is more complex than that of a liquid.

{\vspace*{6cm} \small $^*$: Corresponding author. Email Address:
diambra@if.sc.usp.br. FAX: +55 16 3373 9879.}
\newpage
\baselineskip=1.5\baselineskip
\section{Introduction}

Cell motility has an important role in many biological processes.
On the basis of Abercrombie's work (1970) \cite{aber70} and
subsequent studies, the cell motility cycle can be defined mainly
by five steps: (i) the cells polarize towards a chemo-attractant
signal, eliciting localized actin polymerization
\cite{reve04,welc97}; ii) there is a cell surface rearrangement to
form a protrusion \cite{sant02}; iii) the protrusion makes contact
with the extracellular matrix, or a neighboring cell, to form an
adhesion site \cite{mcne93}; iv) there is an actomyosin-based
contraction, resulting in the development of tension between the
adhesion sites \cite{kobi04}. The last step depends on further
signals that will determine one of two possible physiological
consequences. The first one corresponds to the cell movement. In
this case the fifth step is the detachment of the cell's trailing
edge to reinitiate the cycle again. The exact means by which the
intracellular pathways that control each step of the cell motility
still remain unclear and are the focus of intensive studies.

The locomotion of cells is involved in several physiological
processes, such as the immune response \cite{negu96}, tumor
spreading \cite{chic95}, cell sorting \cite{upad01} and nervous
system development \cite{orou92} and also plays a key role in the
pattern formation during most stages of development
\cite{foty94,rieu98}. Cell movement is highly dependent upon the
environment in which the cells are embedded. Cells moving as part
of a compact cell aggregate interact strongly with each other, so
that cell movement is the result of cell-to-cell interactions as
well as interactions with the surrounding environment. The
locomotory activity of individual cells in media involving low
cell densities is, on the other hand, almost free of cell-to-cell
interactions, but highly dependent on environmental effects. In
both cases the balance of adhesion between cells and the
extracellular environment plays a key role in defining the motion
dynamics of individual cells.

By offering powerful models and measures capable of dealing with
the movement of particles subject to intrinsic and external
effects while accounting for several types of random behavior,
statistical physics represents a unique perspective from which to
approach cell movement \cite{glaz93,momb96,czir98,upad01}. By
using concepts derived from statistical physics, it is possible to
show that cells moving at random (random-walk) are characterized
by simple diffusive processes. This simple type of movement
provides the null hypothesis for the movement of cells in the
absence of intrinsic or extrinsic influences. Previous work on
cell locomotion has focused upon characterizing of the dynamics of
both single cells or groups of cells. These studies have observed
normal diffusive motion and Maxwellian velocity distributions
\cite{momb96,rieu98}. Recently, \cite{upad01} reported anomalous
diffusion associated to velocity distributions within the
framework of the non-extensive thermodynamics in Hydra cells. This
framework could be also useful to analyze the movement of cells
following a chemotactic gradients.

This paper presents a statistical description of the locomotion of
genetically modified human kidney 293T cells and a control cell
line. The specific aim is to uncover the roles of the protein MOCA
in the locomotion of these cells. MOCA (modifier of cell adhesion)
is a protein with 40$\%$ sequence homology to DOCK-180, a protein
that is involved in cell shape and movement via its indirect
interaction with the cytoskeleton \cite{kash00,jali94}. Since the
over-expression of DOCK-180 increases cell migration
\cite{albe00}, it was asked if the expression of MOCA also
influences cell motion. Recent observations suggest that MOCA may
induce cytoskeletal reorganization and changes in cell adhesion by
regulating the activity of Rac1 and N-cadherin
\cite{name04,chen04}. The text provides a series of analytical
methods and resulting insights about the locomotory properties of
these cells.

\section{Materials and Methods}

\subsection{Cells and Time-Lapse}

In the experiments reported here, we used clones of human kidney
293T cells (hereafter named 293T) clones genetically modified to
express MOCA (293-MOCA). The stably transfected MOCA expressing
cell line has been described \cite{chen02}. Cells were cultured on
a laminin surface in a chamber associated with a microscope in
order to capture images of cell movement. Since the phenotypic
characteristics of the cells reflect their culture state,
exponentially growing cultures were used at identical cell
densities of $10^{5}$ cells per 35 mm culture dish (sparse). Both
culture growth medium and the substratum influence cell shape and
movement. Therefore the same experimental conditions were used in
all situations. The maintenance of temperature and medium pH were
accomplished by a heated stage and an enclosed culture chamber in
which a humidified mixture of CO$_2$ and air can be passed. The
chamber was a modified cell culture chamber (Physitemp, Boston
MA), and the experiment was done at 37 degrees Celsius.
Experiments were recorded and translated into digital form by a
time-lapse system at an interval of 35 min per frame of 1022x1280
pixels. The time-lapse system consisted of a digital camera
(Hamamatsu) attached to an inverted microscope (Leitz DMIRB, using
a $16\times$ phase contrast lens), and software for image capture
(OPENLAB by Improvision). With this setup three movies were made
in the experimental conditions indicated above. Two corresponding
to 293-MOCA (denoted by 293-MOCA-A, 293-MOCA-B), and one
corresponding to the control cells 293T (293T-C). Hereafter these
movies will be denoted as experiments A, B and C respectively.
Before being statistically characterized, the movies cells undergo
preprocessing to remove noise, artifacts and enhance the contrast.
Later, each individual frame is extracted from the video sequence
in order to facilitate the cell segmentation and the
reconstruction of cell trajectories. The task of separating cell
from background, (segmentation) was done using a semi-automatic
procedure, where a preliminar segmentation is performed by a
software application and then improved through human intervention.
This software application was developed in Delphi in order to
assisting the operator to mark the soma center of mass for each
moving cell reconstruct the cell trajectory. Fig. 1 shows some of
the trajectories superposed onto the initial acquired frame. Table
1 shows the number of cells followed and the corresponding number
of frames in each experiment.

\subsection{Statistical methods}

With the aim of applying statistical approaches to the study of
cellular locomotory activity, long-term cell migration patterns
were recorded in monolayer cultures. The two-dimensional cell
position was extracted from frames taken each 35 minutes ($\Delta
t =35$) from time-lapse movies by using a computer program. The
trajectory of the cell $i$ is denoted by $\vec{r}_i\left( t\right)
=\left\{ x_i\left( t\right),y_i\left( t\right) \right\}$. The
velocities were estimated by using the mean velocity definition
$\vec{v}_i\left( t\right)= \left( \vec{r}_i\left( t+ \Delta
t\right)-\vec{r}_i\left( t\right)\right)/\Delta t$.

For each experiment we studied: (i) the averaged cell
displacements, (ii) the temporal and spatial correlation function
of velocities, and (iii) the distribution of velocities. The mean
square displacement $\langle z^2 \left( t\right)\rangle$ over $M$
cells in a given experimental setup is given by

\begin{equation}
\langle z^2 \left( t\right)\rangle=M^{-1}\sum _i^{M}\left( \left
(x_i\left( t_0\right)-x_i\left( t\right) \right)^2 +\left
(y_i\left( t_0\right)-y_i\left( t\right) \right)^2 \right)
\end{equation}
which implies $\langle z^2 \left( t\right)\rangle =Dt^{\alpha}$,
where $D$ is an effective diffusivity, $\alpha $ is an exponent
which indicates normal diffusion, like a random walk ($\alpha=1$),
or anomalous diffusion ($\alpha \neq 1$). We measured the
effective constant $D$ and the exponent $\alpha$ by determining
the linear parameters of the plot $\log \langle z^2 \rangle$ vs
$\log t$: the slope is $\alpha$ and the y-axis intercept is $\log
D$. The total displacement of cells was estimated through the
average total path length $\langle L \rangle$ travelled by the
cells in each case. $L$ is the sum of the arc length covered by
the cell, and can be estimated as the sum of the distance between
the subsequent positions along the trajectory. In this way we
avoid extracting a conclusion about cell motion only from the
effective diffusivity $D$, which could introduce wrong conclusions
when one compares normal and anomalous diffusion processes. Low
values of $D$ associated to anomalous diffusion could be not
related to slow locomotion.

When $\alpha=1$, the cells undergo normal diffusion motion like a
random walk. Anomalous diffusion can also be identified by the
following two situations: (i) $\alpha <1$, corresponding to
sub-diffusion, and (ii) $\alpha >1$, which corresponds to
super-diffusion. Anomalous diffusion can be induced by temporal or
spatial correlations \cite{bouc90}. For this reason, we examined
the temporal autocorrelation function and the correlation of
velocities of two pairs of neighbor cells. The former is computed
by averaging the autocorrelation functions over the two components
and over all cells in each experiment
\begin{equation}
C(\tau)=\langle \frac{\sum _{i} (v(t_i) -\overline{v})(
v(t_{i+\tau}) -\overline{v})}{\sqrt{\sum _{i}(v(t_i)
-\overline{v})^2( v(t_{i+\tau}) -\overline{v})^2}} \rangle
\end{equation}

In order to compute the correlation of velocities, we estimate the
velocities of all cell pairs $\{\vec{v}_i,\vec{v}_j\}$ whose
distance are less than $\delta$, i.e., $|\overline{r}_i
-\overline{r}_j|<\delta$, where $\delta $ was taken 25 $\mu m$.
Then the Pearson coefficient $r^2$ and the significance were
computed over the scatter plot $v_i$ versus. $v_j$ including both
velocity components. This correlation can be used to address the
question: Do two neighbor cells travel together more frequently
than cells that are far away from one another?

We also studied the underlying thermodynamics of the motion by
calculating the histogram for the velocity distribution of the
cells, obtained by summing the histograms of individual cells in
each experiment. We fit them to the function $f(v)=av\exp (b
v^2)$, which correspond to the functional form of the Maxwell
distribution function for the velocities of particles in a gas in
two dimensions.

\section{Results}

The following paragraphs details our analysis of the statistical
properties of the cell motion. The top panels of Fig. 2 show the
mean square displacement as a function of time. Table 1 shows the
number of cells tracking in each experiment, the effective
diffusivity constants $D$ and $\alpha $ exponents. In the 293-MOCA
cases the cells execute a random walk associated with a normal
diffusive behavior as indicated by $\alpha \cong 1$, while the
exponent $\alpha $ corresponding to the cell motion in experiment
C is $1.30 \pm 0.1$, suggesting super-diffusion motion, i.e. the
cell movements are less random (coherent motion). Moreover, in
case C, the constant $D$ was 0.7 $\mu $m$^2$/min, substantially
smaller than for the cases A and B, where the $D$ values were 3.4
$\mu $m$^2$/min and 3.3 $\mu $m$^2$/min respectively. Comparing
effective diffusivities may easily produce an incorrect
interpretation since the diffusion process in 293-MOCA cells seems
to be normal, while in control cells seems to be anomalous. For
this reason we also computed the average total length $\langle L
\rangle$ in each case. Table 1 shows the average path length for
each experiment. In case C, the constant $\langle L \rangle$ was
302 $\mu $m, substantially higher than for the cases A and B,
where the $\langle L \rangle$ values were 240 $\mu $m and 250 $\mu
$m respectively, indicating that the cell displacement in the
control case is greater than in 293-MOCA cells. Moreover, bottom
panels of Fig. 2 show the histograms of the speeds and respective
fitting by the Maxwell distribution. The velocity distributions
over the considered time scales ($\Delta t$ = 35 min) follow
essentially the 2D Maxwell-Boltzmann thermodynamics. However, in
all cases the experimental distribution have small peaks around
0.6 $\mu $m/min and 0.9 $\mu $m/min in both 293-MOCA cases, and
around 0.8 $\mu $m/min and 1.1 $\mu $m/min in the 293T cells, as
indicated by the arrows. As a consequence, the distributions
present a longer tail than the corresponding Maxwelian
distribution, which has been observed in other experiments (Fig.
3a in \cite{upad01}) and Fig. 3 in \cite{rieu98}. We should also
note that the velocity distribution associated with 293T cells is
characterized by higher velocities than in the 293-MOCA cases,
which seem to be slower. These observations agree with those
obtained from $\langle L \rangle$ estimation, as shown in Table 1.

In order to avoid cell-cell adhesion effects on the
characterization of the cell motion, we also computed the velocity
distributions for non-interacting cells in each experiment. We
considered all pieces of trajectories where each cell appears
isolated in order to improve the number of events. The ANOVA test
reveals that 293-MOCA cell mean velocity ($\langle v \rangle=0.23
\pm 0.03 \ \mu$m/min) is significatively lower than 293T cells
mean velocity ($\langle v \rangle=0.28 \pm 0.05 \ \mu$m/min), with
a significance level $p<10^{-4}$. The corresponding histograms are
shown in Fig. 3. This result shows that MOCA expression slows cell
locomotion on laminin surfaces relative to control cells
independently of cell-cell contact. It should be notice that the
mean velocities observed for non-interacting cells are smaller
than the mean velocities computed including cells touching each
other (see Table 1).

We have also calculated the temporal and spatial correlation
functions of the velocities. The bottom panel of Fig. 4 shows the
temporal auto-correlation of velocities, which decays very
rapidly, suggesting that temporal correlations are unlikely to be
the cause of the anomalous diffusion observed in experiment C. The
top panel of Fig. 4 depicts the velocities correlogram for all
pairs of cells that are no further than $25\mu$m apart. For
293-MOCA cells (experiments A and B, top panel of Fig. 4), the
velocity components of a cell are correlated to the velocity
component of neighbors cells at the significance level of
$p<10^{-4}$. In contrast, the velocities of two neighboring
control cells are not correlated as shown at right of the bottom
panels of Fig. 4. This means that 293-MOCA cells travel together
more frequently than 293T cells. These data show that the MOCA
protein may be involved with cell-cell in adhesion and that it
also influences cell movement. This conclusion is also supported
by Fig. 5, which depicts the mean distance between cells which
were no further apart than $25\mu$m in the initial frame as a
function of time. It is clear from this result that the 293-MOCA
cells remain closer together than the wild type.

\section{Discussion and Conclusion}

Cell movement requires a series of highly coordinated events that
are powered by the actin cytoskeleton and regulated by a complex
group of kinases and phosphatases (for reviews see
\cite{mogi02}).
MOCA is an abundant protein that is found in the hippocampus and
other cortical areas of the brain \cite{kash00}.
Recently has been demonstrated that the expression of MOCA is
required for neurite outgrowth in both PC12 cells and central
nervous system neurons \cite{chen04}. The above data show that the
expression of MOCA in 293T cells, that normally lack this protein,
leads to two changes: i) a decrease the cell motion and ii) an
increase the co-migration with neighboring cells. The decrement of
the overall displacement in isolated 293-MOCA cells (Fig. 3)
suggests that first conclusion could not to be a consequence of an
increase in MOCA induced cell-cell adhesion. The second set of
results are, however consistent with our data demonstrating that
MOCA increases N-cadherin-mediated cell-cell adhesion
\cite{chen04}. In contrast, the over expression of DOCK-180 leads
to increased cell migration \cite{albe00}, establishing that the
function of the two proteins are distinct.

The results presented here establish that MOCA protein expression
modifies two important aspects of cell movement. The measurements
clearly indicate that MOCA tends, by enhancing of joint motion, to
inhibit the super-diffusion behavior observed in the wild type
cells. The broader distributions of velocities for that case could
be the reason for super-diffusion. In fact, wild type cells have
higher velocity than the modified cells. Furthermore, the spatial
correlations of the cell velocities show that two neighboring
293-MOCA cells travel together more frequently than two control
cells, and that their velocities are quite correlated. These facts
are compatible with the hypothesis that the cells move faster in a
less cohesive environment.

We also determined the temporal autocorrelation of the velocity.
In all cases we obtained a single narrow peak at null lag,
reasonably approximating a delta function. The correlation time
for the velocity was shorter than the measurement interval used in
the experiments (35 min), agreeing with the hypothesis of
time-uncorrelated velocities in Brownian motion. Preliminary
results on shorter time scales (5 min) do not suggest
correlations.

Finally, the observation that the experimental distribution tails
are longer than those predicted by Maxwell-Boltzmann
thermodynamics in all cases studied here, agrees with results in
previous studies \cite{upad01,rieu98}. This fact could be a
statistical indication that cell movement is more complex than
that of a liquid.

\section*{Acknowledgements}

Luciano da F. Costa is grateful to FAPESP (proc. 99/12765-2), CNPq
(proc. 301422/92-3) and Human Frontier Science Program for
financial support. Luis Diambra thanks the Human Frontier Science
Program for his post-doc grant.

\newpage

\section*{Figure Legends}

\noindent Figure 1: Some cell trajectories superposed onto the
initial acquired frame corresponding to experiment A. Inset: A
sample trajectory enlarge containing 13 successive positions of
cell $i$.

\noindent Figure 2: Top panels: $\log(\langle z^2 \rangle)$ versus
$\log(t)$ plots for each experiment. Bottom panels: Histogram of
the 2D cell velocity distribution. The solid curve is a fit by the
function $f(v)=av\exp (b v^2)$ (2D Maxwellian distribution).
Arrows are indicating deviation from Maxwellian distributions. The
left and center panels correspond to experiments A and B
respectively, and the right panels correspond to experiment C.

\noindent Figure 3: Histogram of velocity for noninteracting
cells. The dashed bars correspond to cell velocities from A and B
experiment and solid grey bars correspond to cell velocities from
experiment C.

\noindent Figure 4: Top panels: The paired velocity correlogram
including both velocity components of all cell pairs whose
distance are smaller than 25$\mu$m in each experiment. In each
case the correlation coefficient as well its significance is
displayed. Bottom panels: The auto-correlation function $C(t)$, of
the cell velocities. The left and center panels correspond to
experiments A and B respectively, and the right panels correspond
to experiment C.

\noindent Figure 5: Mean distance between cells whose initial
distance were smaller than 25$\mu$m as a function of time. A and B
case in solid and dashed line respectively; C in dotted line.

\newpage
\section*{Tables}
\begin{table}[ht]
\begin{tabular}{|c||c|c|c||c|c|c|}
  \hline
  Experiment & $\#$ of frames & $\#$ of cells & $\langle v \rangle$ [$\mu $m/min] & $\langle L \rangle$ [$\mu $m] & $D$ [$\mu {\rm m}^2$/min] & $\alpha$ \\ \hline
   \hline

  A & 31 & 53 & 0.25 & 250 & 3.379 & 1.01$\pm$0.14\\
  B & 41 & 43 & 0.24 & 240 & 3.265 & 1.05$\pm$0.11\\
  C & 37 & 33 & 0.30 & 302 & 0.703 & 1.30$\pm$0.10\\ \hline
\end{tabular}
\caption{Some important features of the data investigated and
$\langle L \rangle$, $D$ and the exponent $\alpha$.}
\end{table}

\newpage

\begin{figure}[ht]
\includegraphics[width=15cm]{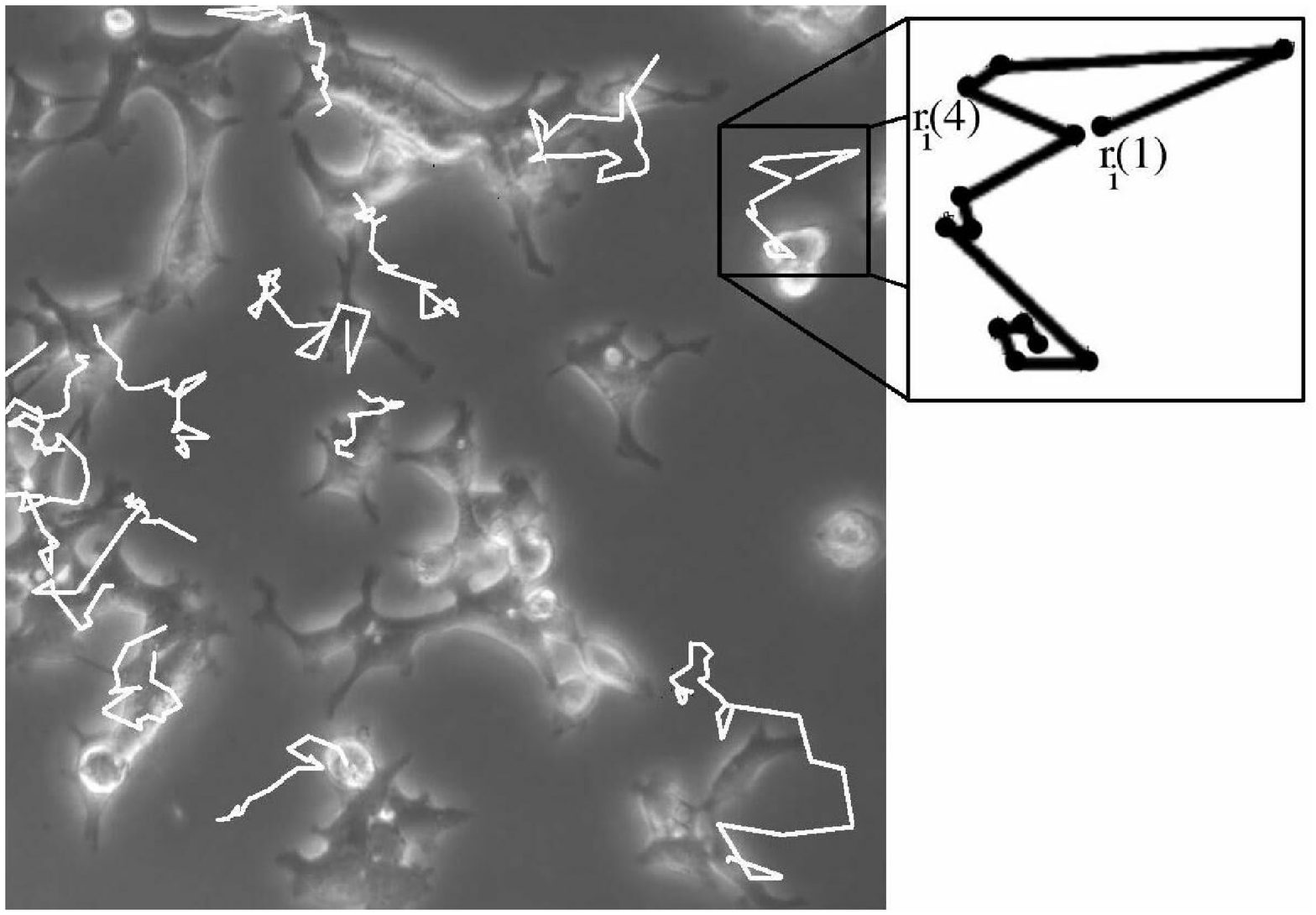}
\caption{}
\end{figure}

\begin{figure}[ht]
\includegraphics[width=15cm]{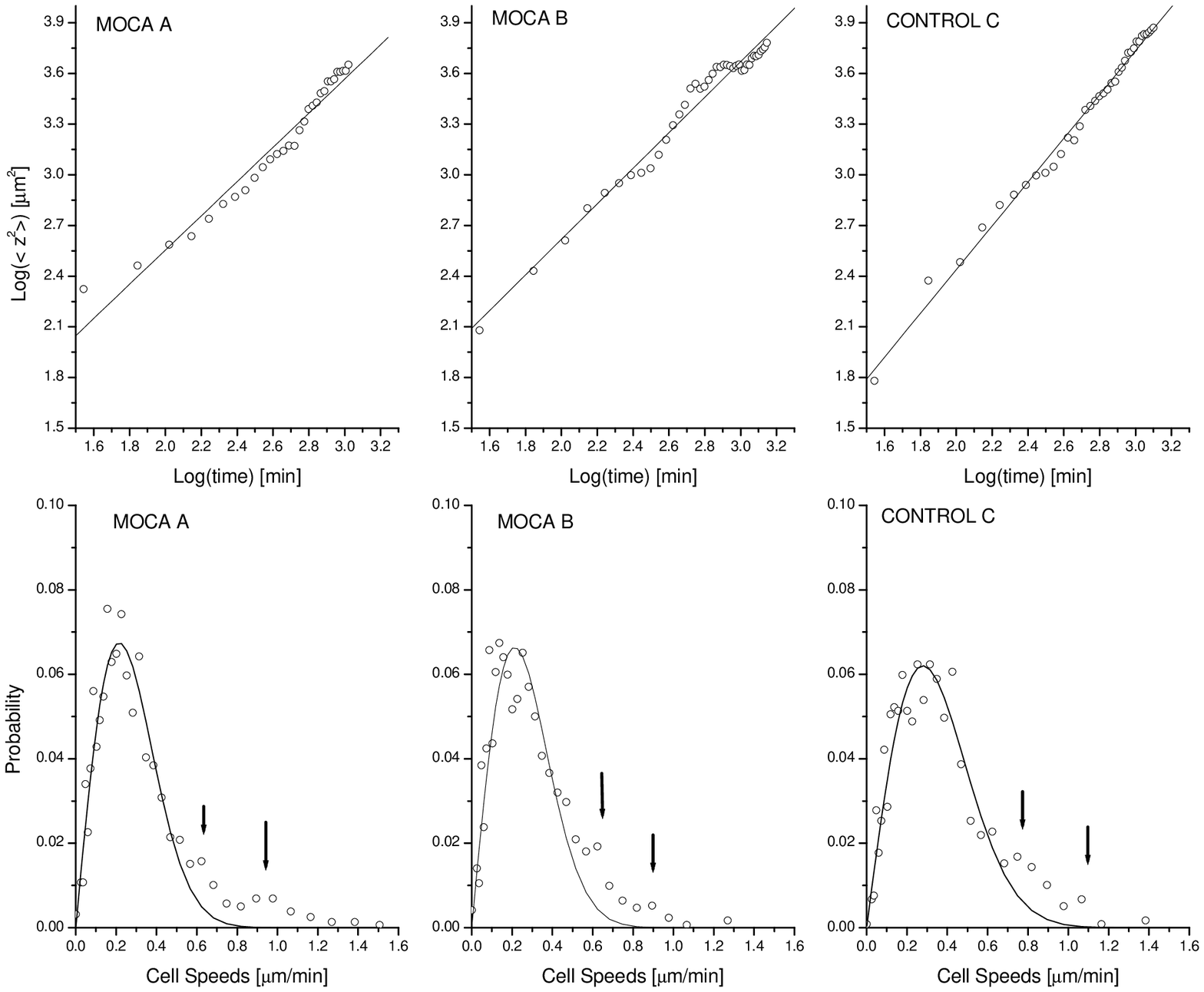}
\caption{}
\end{figure}

\begin{figure}[ht]
\includegraphics[width=15cm]{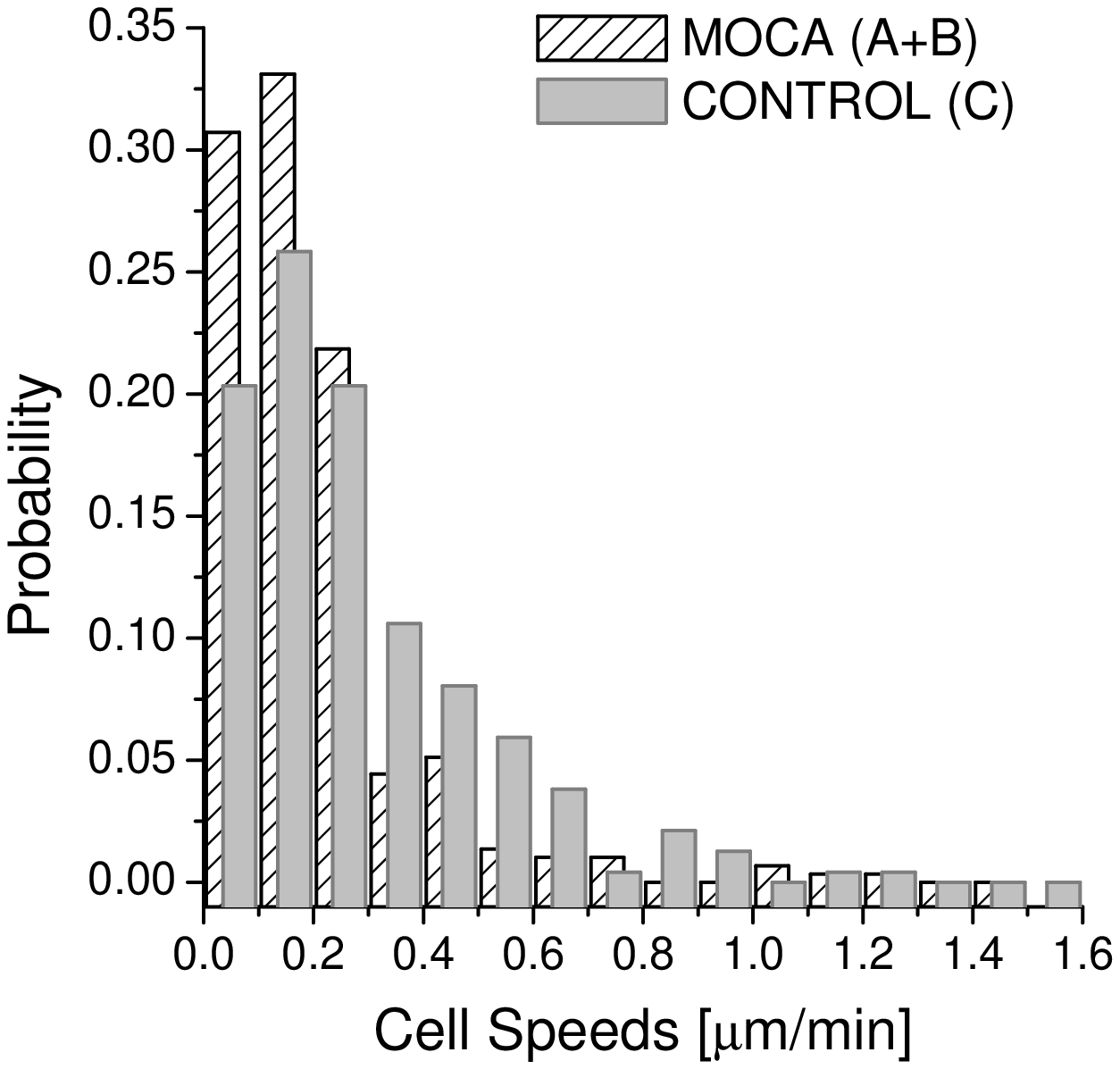}
\caption{}
\end{figure}

\begin{figure}[ht]
\includegraphics[width=17cm]{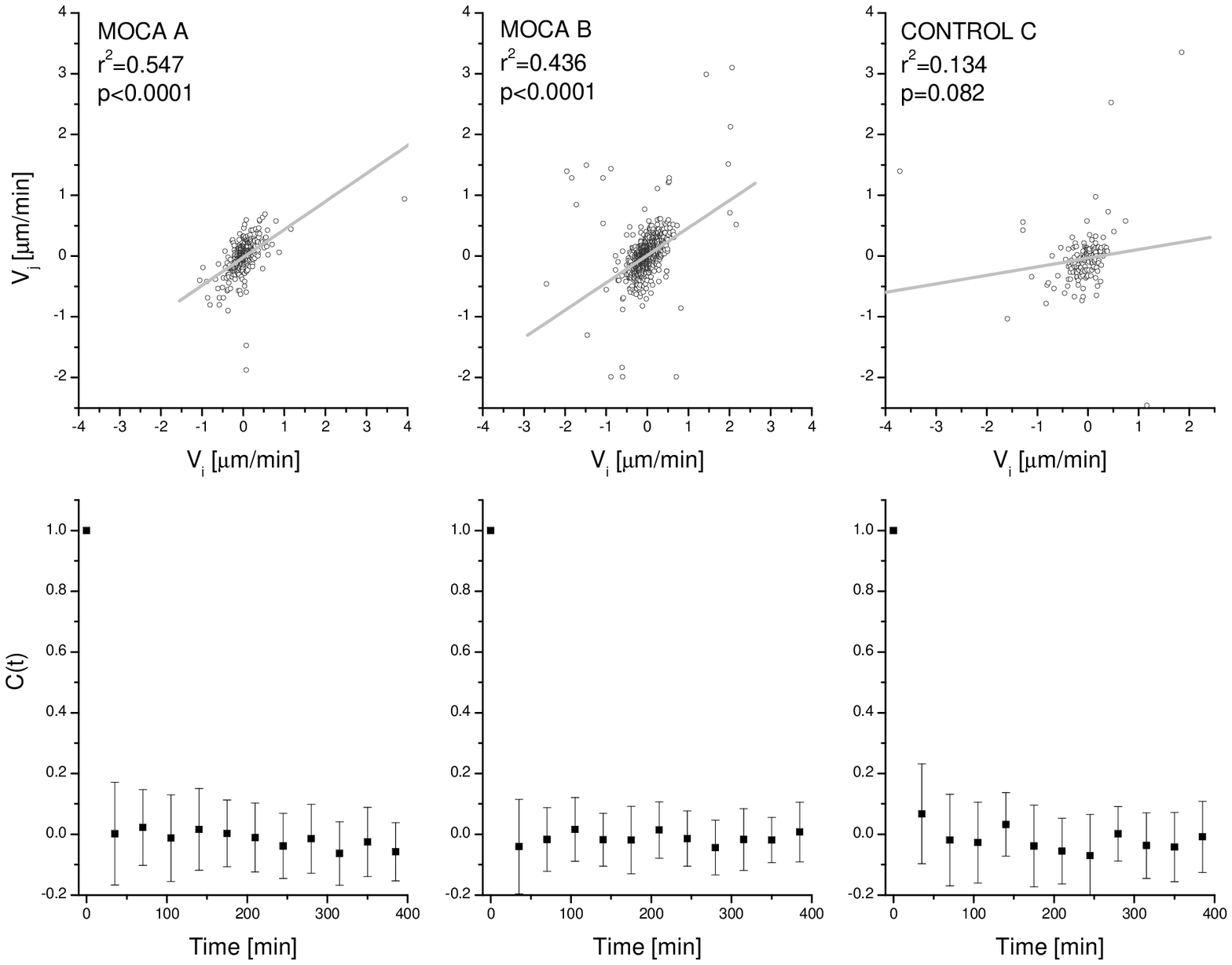}
\caption{}
\end{figure}

\begin{figure}[ht]
\includegraphics[width=15cm]{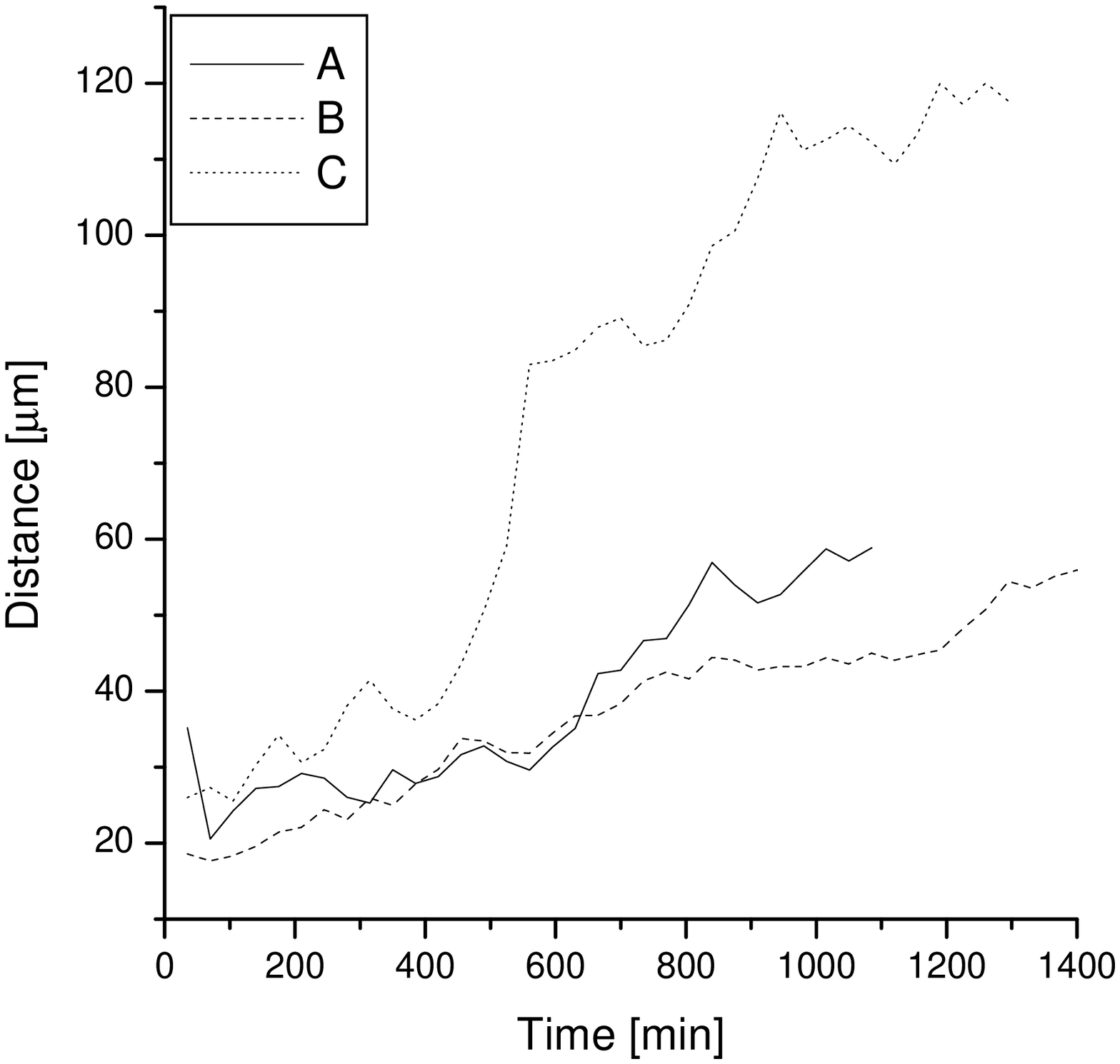}
\caption{}
\end{figure}
\end{document}